# Fluorescence lifetime modification in Eu:Lu$_2$O$_3$ nanoparticles in the presence of silver nanoparticles

*Haoyan Wei, Zachary Cleary, Sang Park, Keerthesinghe Senevirathne, Hergen Eilers*

Applied Sciences Laboratory, Institute for Shock Physics, Washington State University

Spokane, WA 99210-1495

ABSTRACT

Europium-doped lutetium-oxide (Eu:Lu$_2$O$_3$) nanoparticles were synthesized using a combustion technique and a co-precipitation technique, and their properties were compared. Surface-modification utilizing small silane molecules and long chain polymers were explored to de-agglomerate and disperse the particles. Structural, morphological and optical properties were characterized with x-ray diffraction, scanning and transmission electron microscopy, and laser spectroscopy respectively to evaluate these materials. The luminescent behaviors were compared between the pristine and modified Eu:Lu$_2$O$_3$ nanoparticles to study the influence of surface ligands on emission properties. Subsequently, the Eu:Lu$_2$O$_3$ nanoparticles were placed on top of a thin film consisting of silver nanoparticles and combined with silver nanoparticles and dispersed in a polymer matrix. The presence of the silver nanoparticles led to a reduction of the fluorescence lifetime of 12-14%.

# 1. INTRODUCTION

The threat of nuclear and radiological attacks by terrorists has placed additional emphasis on the development of new scintillator materials used for security screening in border and port controls [1-6]. In general, it is desired to have scintillator materials with a high density and a high light output (photons per MeV). In recent years, there has been a growing interest in Eu:$Lu_2O_3$ materials because of their merits of a very high density of 9.4 g/cm$^3$ and resultant high stopping power for ionizing radiations. As a result, they can be physically engineered into thinner screens with superior spatial resolution. The primary emission band of Eu:$Lu_2O_3$ is *ca*. 610 nm, which makes this material highly attractive for use as X-ray phosphors in digital radiography since most CCD detectors are most sensitive to the red light. Counting applications in surveillance, as opposed to imaging applications, also require that the scintillator material has a short luminescence lifetime, ideally in the nanosecond-range. However, the luminescence lifetime of Eu:$Lu_2O_3$ is about 1.5~2 ms, relatively long for counting applications. Different dopants with a shorter luminescence lifetime that are suitable for the $Lu_2O_3$ host have not been identified, yet.

Alternatively, radiative-decay engineering could be a viable approach to overcome this limitation [7-15]. In this approach, the luminescence rates are modified through the close proximity of metallic nanostructures which alter the portion of energy into the radiative and non-radiative pathways. This approach has been successfully applied to the fields of bio-medical imaging and bio-sensing [11-13]. The fluorescence lifetime ($\tau_0$) and quantum efficiency ($Q_0$) of a material with the presence of a nearby metal can be described as:

$$\tau_0 = \frac{1}{(\Gamma + \Gamma_m + k_{nr})} \qquad (1)$$

$$Q_0 = \frac{\Gamma + \Gamma_m}{(\Gamma + \Gamma_m + k_{nr})} \qquad (2)$$

Where $\Gamma$ is the radiative decay rate, $k_{nr}$ is the non-radiative decay rate and subscript *m* denotes metal. As indicated in the above equations, it is possible to simultaneously decrease the decay time and increase the luminescence intensity through the introduction of a metallic luminescent decay rate ($\Gamma_m$). This manipulation is possible because the metallic luminescent decay rate and the metallic non-radiative decay rate operate on different length scales and can thus be de-coupled.

The enhancement of luminescence from $Eu^{3+}$ ions by metallic particles has been studied previously on host materials of glass and gels [16-18]. Recently the use of different host matrix $Lu_2O_3$ containing silver particles was also reported [19]. However, the effect on the fluorescence life time is less investigated. In these systems, the metal particles are incorporated through the direct reduction of metal cations from premixed solution with the host materials. The distance between metal particles and $Eu^{3+}$ activator ions are not controlled. Since the fluorescence response is the competition between the enhancement from the local field around metals and non-radiative relaxation due to damping of fluorophore-oscillating dipoles, the spacing between metal particles and fluorophores is crucial for the performance optimization.

Maximizing the interaction and fine tuning the distance between metal particles and fluorophores require the synthesis of nanophase $Eu:Lu_2O_3$ and their surface functionalization to subsequently control the placement of metallic nanostructures in close proximity. Since the luminescent emission is sensitive to the boundary and interface conditions, the effect of surface ligands has yet to be investigated. In this contribution, we compare two popular synthesis routes of $Eu:Lu_2O_3$ nanoparticles. Initial approaches for surface modification against agglomeration were explored on the synthesized materials. The influence of surface coupling agents and metal particles on optical properties, including reduction in fluorescence lifetime of this material, is discussed.

## 2. EXPERIMENTAL

$Eu:Lu_2O_3$ nanocrystals were synthesized using two methods, a combustion method [20-22] and a co-precipitation method [23]. Europium nitrate pentahydrate ($Eu(NO_3)_3 \cdot 5H_2O$ 99.999%) and lutetium nitrate hydrate ($Lu(NO_3)_3 \cdot 5H_2O$ 99.999%) were obtained from MV Laboratories, Inc. in NJ, USA. For the combustion method, these materials were dissolved in deionized (DI) $H_2O$ with a designated 5 at% Eu concentration and stirred for 10 minutes at room temperature. Glycine (Sigma-Aldrich, USA) was adopted as the fuel and added to the solution with a stoichiometric oxidizer/fuel molar ratio of 0.6, and stirred for 20 more minutes at room temperature. While continuing to stir, the solution was heated to 80°C until the sample had a gel-like consistency. Subsequently, the sample was heated to 220ºC to initiate the combustion reaction. After 2 hours the sample was heated to 800ºC and kept there for 2 hours to complete the calcination process.

For the co-precipitation method, a europium nitrate pentahydrate and lutetium nitrate hydrate mixture was dissolved in DI $H_2O$ at the same preset dopant level of 5 at% Eu concentration. Under mild agitation, $NH_4HCO_3$/$NH_3 \cdot H_2O$ precipitant was then drop-wise added to the $Lu(NO_3)_3$/$Eu(NO_3)_3$ solution until the precipitation started to occur. The pH value was controlled between 8 and 9. Subsequently, this precursor suspension of basic carbonate $Lu(OH)_x(CO_3)_y$ was vacuum filtered and rinsed with DI $H_2O$ four times. The sample was dried at 110°C for 24 hours before being calcined at either 850°C or 1000ºC for 2 hours.

Three methods of surface functionalization were explored to disperse the as-synthesized nanopowders for subsequent manipulation. The first method uses neutral chlorosilane molecules to render the Eu:$Lu_2O_3$ surface hydrophobic. Eu:$Lu_2O_3$ powders were dispersed in acetone under mild agitation. Subsequently, 1-3 vol% dimethyldicholorosilane, $(CH_3)_2SiCl_2$, (DMDCS) (>99%, Sigma-Aldrich, USA) was drop-wise added to the solution, which was then magnetically stirred at the boiling temperature of acetone (56°C) under flowing $N_2$ for 180 minutes. The solution was centrifuged and the particles that remained were washed with acetone and dried in air. These particles were then dispersed in dimethyl sulfoxide (DMSO) (J.T. Baker), which resulted in a clear dispersion.

The second surface-functionalization method used alkoxysilane molecules with a hydrophilic amino groups at the end – 3-aminopropyltriethoxysilane, $NH_2(CH_2)_3Si(EtO)_3$, (3-APTES) (>98%, Alfa Aesar). The same dispersion procedure as in the first technique was followed, except that the concentration of 3-APTES in the mixture was 5 vol%. The modified particles were then dispersed in DI water.

The third approach utilized polymerization of methacrylic acid (MAA) to form an overcoat layer of polymethacrylic acid (PMAA). It was designed to exploit the benefit of steric hindrance caused by the polymer adsorbed on nanoparticle surfaces and the electrostatic repulsion between nanoparticles due to charged carboxylic acid endgroups. This approach expanded the tunable range of overcoat layer thicknesses. The surface functionalization process started by gently grinding 45.1 mg Eu:Lu$_2$O$_3$ nanoparticles in acetone using an alumina mortar and pestle. This treatment can effectively break down a number of coarse and hard agglomerates that are present in the as-synthesized Eu:Lu$_2$O$_3$ powder. The dried Eu:Lu$_2$O$_3$ nanoparticles were added to 100 mL DI water which was then sonicated for five minutes. The solution was heated to 75°C under a N$_2$ atmosphere while being vigorously agitated. At 75°C, 24 μL MAA monomers were added to the solution, and after 45 minutes 3.5 mg potassium persulfate (K$_2$S$_2$O$_8$) was added to initiate the polymerization of MAA. The solution was subsequently stirred for another 120 minutes. Finally, the pH value of the solution was adjusted to about 10 to form a stable dispersion.

The as-synthesized nanoparticles were characterized by x-ray diffraction (XRD) using the Cu K$_\alpha$ radiation from a PANalytical X-Pert Pro Diffractometer. Microstructural information was obtained on the as-synthesized and the functionalized nanoparticles using an FEI Sirion 200 field emission scanning electron microscope (FESEM) and a JEOL JEM 1200 EX transmission electron microscope (TEM). The particle size distribution of dispersed nanoparticles was characterized by dynamic light scattering (DLS) using a NICOMP Particle Size Analyzer. Optical measurements were performed using a Continuum Q-switched Nd:YAG/OPO laser system (8 ns pulse lengths and tunable from 400 nm to 2400 nm) and a 750 mm Acton monochromator with a CCD array for fluorescence and excitation measurements and a PMT for fluorescence lifetime measurements.

## 3. RESULTS AND DISCUSSION

Figure 1 shows a comparison of three XRD scans of the as-synthesized Eu:Lu$_2$O$_3$ nanopowders. The XRD scans confirm that both synthesis techniques lead to the formation of cubic Lu$_2$O$_3$. Since both samples were annealed, the peak broadening is assumed to be arising only from the size effect of the crystallites, without significant micro strain. The volume averaged grain size (D) is estimated using the Scherrer equation $D = 0.9\lambda/\beta\cos\theta$, where $\lambda$ is the X-ray wavelength, $\theta$ is the diffraction angle, and $\beta$ is the full-width at half-maximum (FWHM) in radians. The peak broadening was corrected for the instrumental broadening effect by measuring a Si wafer as a standard sample. The true peak broadening is then obtained using the following relationship $\beta^2 = \beta_{exp}^2 - \beta_{inst}^2$ where $\beta$ is the true peak broadening, $\beta_{exp}$ is the FWHM of the experimental observation, and $\beta_{inst}$ is the FWHM of the standard sample. The XRD results show that the average diameter of the particles synthesized using the co-precipitation method was about 20 nm and 28 nm, respectively for calcination temperatures of 850ºC and 1000ºC. The average diameter of the particles synthesized using the combustion method was about 42 nm.

Figure 2 shows an SEM image of nanophase Eu:Lu$_2$O$_3$ synthesized via the combustion technique. The rapid release of gas by-products during the synthesis leads to a very large volume change and results in sponge-like fluffy powders with large air inclusions. In addition, the combustion reaction kinetics is difficult to control in a precise manner. The TEM image of this sample, shown in Figure 2a, indicates that the combustion synthesized Eu:Lu$_2$O$_3$ nanoparticles have experienced partial sintering (hard agglomeration).

Figure 2 also shows an SEM and TEM image of nanophase Eu:Lu$_2$O$_3$ synthesized via the co-precipitation technique. These Eu:Lu$_2$O$_3$ nanoparticles appear to be more loosely bound, which is due to the gradual release of CO$_2$ during the decomposition of the carbonate precursor. However, in both cases severe aggregation is visible, which is induced by the large free surface area of the nanoparticles. Further processing of Eu:Lu$_2$O$_3$ nanoparticle assemblies requires their dispersion in order to maximize their interaction with nearby Ag nanoparticles. The results of several dispersion approaches are described later in this report.

The particle size distribution, obtained from electron microscopy imaging, for combustion and co-precipitation (850ºC) is shown in Figure 3. Both sets of data fit very well with lognormal distribution profiles. The arithmetic mean diameter of the nanoparticles synthesized via co-precipitation (850ºC) is about 19 nm, in good agreement with the Scherrer size determined by XRD analysis. Combustion synthesis results in a much broader distribution of particle sizes than co-precipitation. Due to the much broader size distribution and the presence of some very large particles (over 150 nm), a relative larger discrepancy exists between the number averaged grain size determined via TEM and the volume averaged grain size determined via XRD.

Figure 4 shows the fluorescence spectra of Eu:Lu$_2$O$_3$, synthesized via co-precipitation and combustion. The samples were excited at 467.75 nm. The emission characteristics for these two samples were nearly identical, and were dominated by an emission peak at 611.32 nm. Other notable peaks were observed at 580.5 nm, 587.2 nm, 593.0 nm, 600.3 nm, and 632.1 nm. These peaks were identical for particles synthesized using co-precipitation and combustion techniques and could be assigned to transitions from $^5D_0$ to $^7F_J$ as indicated in Figure 4 [24].

The fluorescence lifetime (611.32 nm) for the co-precipitation synthesized Eu:Lu$_2$O$_3$ nanoparticles was observed to be a single exponential decay of 2.3 ms. The single exponential fit of the fluorescence lifetime for the combustion synthesized Eu:Lu$_2$O$_3$ nanoparticles yielded 1.62 ms. One explanation for the longer fluorescence lifetime of the Eu:Lu$_2$O$_3$ nanoparticles synthesized by co-precipitation is that these particles may have a slightly larger degree of crystallinity than the combustion synthesized nanoparticles. Unlike the fast chemical process involved in combustion synthesis, the co-precipitation process appeared to have a better distribution of the dopant ions within the host material, leading to more uniform crystal structures.

To break down the large agglomerates of nanopowders for subsequent manipulation, three distinct approaches were explored to form stable dispersions in different environments, including organic and aqueous media. The coupling mechanism for all three approaches is based on the condensation reaction between the hydroxyl groups on the oxide surface and head groups (-Cl, -OC$_2$H$_5$, -COOH) of the coupling agents. The SEM images in Figure 5 show that the silane-based (DMDCS and 3-APTES) surface functionalization and the PMAA-based surface functionalization of Eu:Lu$_2$O$_3$ nanoparticles resulted in a different level of agglomeration reduction. The DMDCS and PMAA surface functionalized nanoparticles exhibit a significantly better dispersion than the 3-APTES coated ones. The dispersion liquid appears transparent for the prior two approaches, while the 3-APTES dispersion is subject to flocculation. The separation of the nanoparticles can be clearly observed in the images, with the presence of a black gap between the particles indicating their dispersion. The 3-APTES-based surface functionalization led to the formation of large particles (~1 µm) that formed agglomerates. Most likely, these particles contain Eu:Lu$_2$O$_3$ nanoparticles.

The size distribution of the DMDCS and PMAA modified Eu:Lu$_2$O$_3$ nanoparticles in dispersion were measured with dynamic light scattering (DLS) as shown in Figure 6. It further confirms the important role of DMDCS and PMAA in the separation and subsequent stabilization of Eu:Lu$_2$O$_3$ nanoparticles. The proposed chemical grafting mechanisms are schematically illustrated in Figure 5. For organo-silanes, their head groups (-Cl or –OC$_2$H$_5$) hydrolyze to hydroxyl moieties under the presence of water. Although the amount of moisture present in solvent and particles was minute, they were effectively concentrated to the oxide surface which behaves as drying agent. This water was required to activate the silane condensation with the hydroxylized oxide surface. However, a strong side reaction (inter-silane condensation) could occur in parallel.

The presence of three head groups in 3-APTES resulted in excess self-polymerization, forming a very thick overcoat layer and severe crosslink between coated particles as indicated in Figure 5b. In contrast, DMDCS has fewer head groups and stronger interactions with hydroxylized oxides [25]. That means DMDCS reacts with not only a single hydroxyl group but also its neighbor (Figure 5d). This reaction primarily forms a monolayer on the oxide surface with two chemical bonds between Si and oxides [26]. The coated particles are separated in the liquid due to the physical presence of a silane overcoat (steric hindrance). A third type of coupling agent – a polymer – is used to form dispersions in aqueous media and extend the potential tunable range of the coating thickness via controlled polymerization. The PMAA anchored on Eu:Lu$_2$O$_3$ through carboxyl groups and the large polymer backbone fulfill the steric hindrance. On the outermost layer, un-grafted carboxyl groups oriented outwards, lead to hydrophilic behavior. This hydrophilicity enables a stable dispersion in an aqueous medium. In addition, electrostatic repelling due to negatively charged carboxyl groups contributes further to the dispersion.

While in dispersion, the surface-functionalized samples showed very little or dramatically reduced fluorescence, with the PMAA surface-functionalized material yielding the highest level of luminescence amongst these dispersions. This reduction in fluorescence is most likely due to energy transfer from Eu-ions near the Eu:Lu$_2$O$_3$ nanoparticle surface to solvent molecules. However, after centrifugation and removal of the solvent, the fluorescence signal recovered again.

Figure 7 shows the fluorescence spectra of DMDCS surface-functionalized co-precipitated Eu:Lu$_2$O$_3$ nanoparticles and DMDCS surface-functionalized combustion-synthesized Eu:Lu$_2$O$_3$ nanoparticles. The signal-to-noise ratio is greatly reduced in contrast to that of the as-synthesized nanopowders. The fluorescence signal from DMDCS surface-functionalized co-precipitated Eu:Lu$_2$O$_3$ nanoparticles was so weak that no fluorescence lifetime data acquisition was attainable. This lower signal may be due to the smaller size of the nanoparticles synthesized via co-precipitation resulting in a larger surface/volume ratio, thus affecting the optical properties of co-precipitation-synthesized materials in a larger degree than combustion-synthesized materials. The short fluorescence lifetime (1.25 ms for combustion-synthesized Eu:Lu$_2$O$_3$) and the weak fluorescence signals indicate that the interaction with DMDCS results in significant non-radiative decay channels.

Figure 8 shows the fluorescence spectra of centrifuged 3-APTES surface functionalized co-precipitated Eu:Lu$_2$O$_3$ nanoparticles ($\tau_0$=2.2 ms) and of PMAA coated co-precipitated Eu:Lu$_2$O$_3$ nanoparticles (in dispersion). The present measurement on the fluorescence lifetime of PMAA coated co-precipitated Eu:Lu$_2$O$_3$ nanoparticles did not yield conclusive results. As the pH and the amount of salt changes, so does the fluorescence lifetime. These fluorescence lifetime data require further investigation before conclusions can be drawn regarding the PMAA surface functionalization.

However, it appears as though 3-APTES and PMAA affect the optical properties of Eu:Lu$_2$O$_3$ less than the DMDCS surface functionalization process. The fluorescence signal from 3-APTES surface functionalized co-precipitated Eu:Lu$_2$O$_3$ nanoparticles showed the best S/N ratio, while the fluorescence signal from the PMAA coated co-precipitated Eu:Lu$_2$O$_3$ nanoparticles had a weak S/N ratio. Energy transfer from Eu ions near the surface of the Eu:Lu$_2$O$_3$ nanoparticles appears to be the least affected for 3-APTES. This may be due to the lesser degree of reaction between 3-APTES and the Eu:Lu$_2$O$_3$ nanoparticles, as indicated in Figure 5.

To study the effectiveness of metal-enhanced fluorescence decay rate we used two different approaches. The first approach is a thin-film structure, in which we deposited by thermal evaporation a thin film consisting of silver nanoparticles onto a microscope slide [27]. Subsequently, we dispersed combustion-synthesized Eu:Lu$_2$O$_3$ nanoparticles in a solvent and placed a drop of this dispersion on top of the silver layer. Figure 9 shows a design schematic and SEM images of the Ag nanoparticles and Eu:Lu$_2$O$_3$ nanoparticles after placement on top of the silver nanoparticles.

Figure 10 shows a comparison of the fluorescence spectra and fluorescence lifetimes for Eu:Lu$_2$O$_3$ nanoparticles and Eu:Lu$_2$O$_3$ nanoparticles on top of silver nanoparticles. The fluorescence spectra for the two samples appear to be identical, with no apparent differences. However, the fluorescence lifetime for Eu:Lu$_2$O$_3$ nanoparticles on top of silver nanoparticles is 14% shorter than the fluorescence lifetime of Eu:Lu$_2$O$_3$ nanoparticles without silver. This effect was expected from equation 1 and is consistent with the concept of metal-enhanced fluorescence or radiative decay engineering.

For the second approach, we mixed combustion-synthesized Eu:Lu$_2$O$_3$ nanoparticles along with silver nanoparticles into a polystyrene matrix without the use of any surface modifiers, resulting in a brown-colored bulk sample. Again, the fluorescence spectra appear to be identical, while the fluorescence lifetime is shortened – by 12% in this case. Using equations 1 and 2, the measured lifetimes of 1.62 ms and 1.39 ms in Figure 10, as well as a quantum efficiency of 85% [28], one can calculate $\Gamma$=525 s$^{-1}$, $k_{nr}$=92.6 s$^{-1}$, $\Gamma_m$=102 s$^{-1}$, and $Q_0$=87.13%. Because Eu:Lu$_2$O$_3$ starts out with a relatively high quantum efficiency of 85%, the observed changes are relatively small. However, the initial observations confirm that it is possible to reduce the fluorescence lifetime of scintillator materials. Nevertheless, the implementation of this approach requires further investigations on the improvement of surface functionalization and dispersion of these materials, the influence of surface ligands on optical responses, as well as optimization of interparticle spacing between colloidal metallic nanoparticles and Eu:Lu$_2$O$_3$ nanoparticles.

## 4. CONCLUSION

Nanoparticles of the scintillator material Eu:$Lu_2O_3$ were successfully synthesized using combustion and co-precipitation. The co-precipitation method yielded particles with a diameter of about 25 nm and the combustion technique yielded particles with a diameter of about 40 nm. Three surface modification techniques were explored to improve the dispersion of the nanoparticles, and their effects on the fluorescent emission were studied. DMDCS surface functionalization resulted in significantly better dispersion than 3-APTES and PMAA based approaches. However, this approach significantly reduced the fluorescence of the material. As expected, the combination of the Eu:$Lu_2O_3$ nanoparticles with silver nanoparticles led to reductions of the fluorescence lifetime of 12-14% in the present design. Further tailoring of the optical properties depends on the optimization of the separation between metal and Eu:$Lu_2O_3$ nanoparticles through improved dispersion and surface functionalization.

## ACKNOWLEDGEMENT

This work was supported by DHS/DNDO Grant 2008DN-077ARI004-02, NSF Grant CBET-0735911, and ONR Grant N00014-03-1-0247.

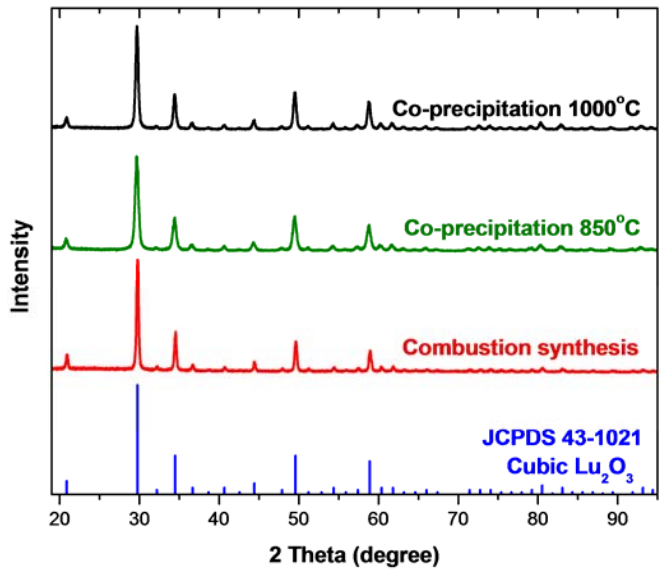

Figure 1. XRD scans of three sets of nanophase Eu:Lu$_2$O$_3$

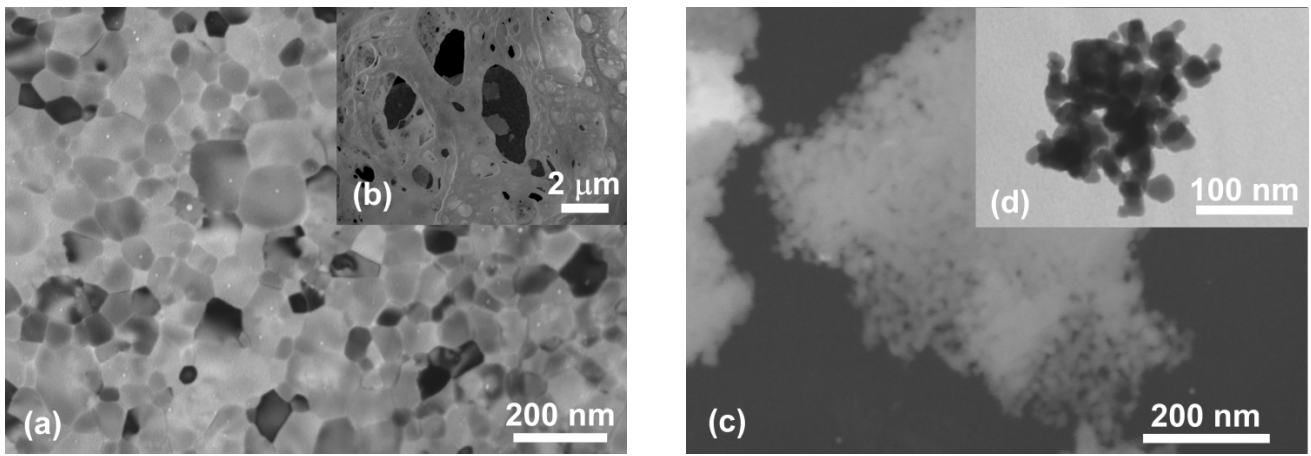

Figure 2. Electron microscope images of Eu:Lu$_2$O$_3$ nanoparticles synthesized via combustion followed by 800°C - 2 hour calcination (left) and co-precipitation followed by 850°C - 2 hour calcination (right). (b) and (c) are SEM and (a) and (d) are TEM images.

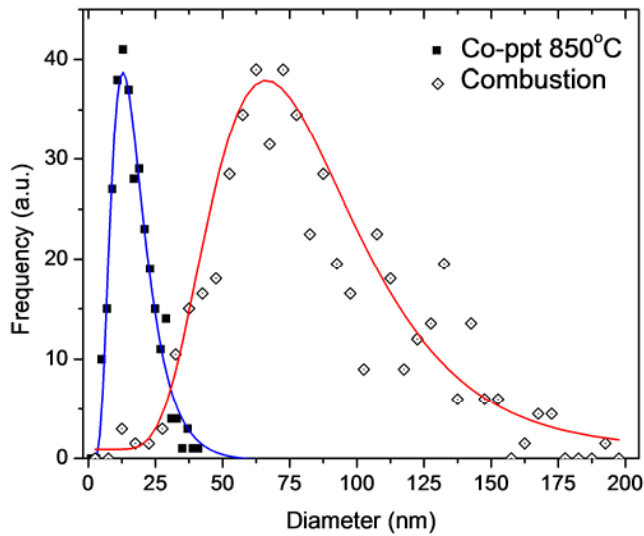 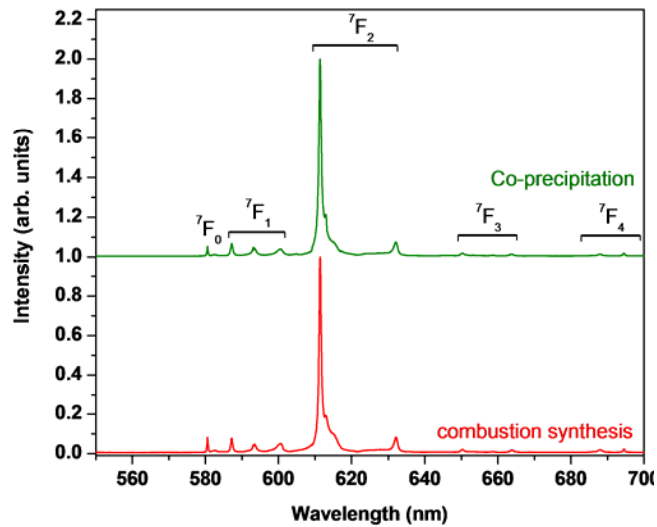

Figure 3. Particle size distributions for combustion-synthesized and co-precipitation-synthesized Eu:Lu$_2$O$_3$ nanoparticles and superimposed log-normal curve fittings. The data was determined from the analysis of electron microscope images using ImageJ software. The combustion synthesized nanopowders have a significantly wider size distribution as well as much larger particles than the co-precipitation synthesized nanopowders.

Figure 4. Fluorescence spectra of Eu:Lu$_2$O$_3$ nanoparticles synthesized via combustion (bottom) and co-precipitation (top).

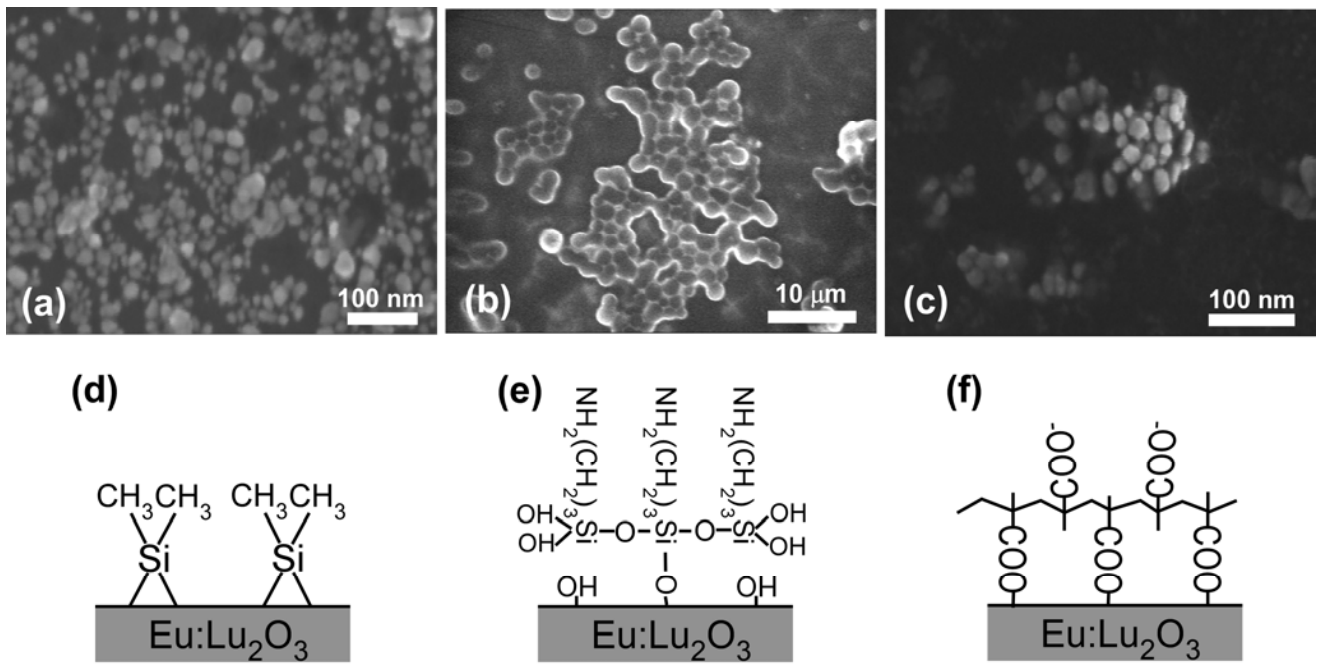

Figure 5. SEM images of DMDCS-DMSO-based surface functionalization (left), 3-APTES-based surface functionalization (center), and PMAA-based surface functionalization (right) of Eu:Lu$_2$O$_3$ nanoparticles synthesized by co-precipitation.

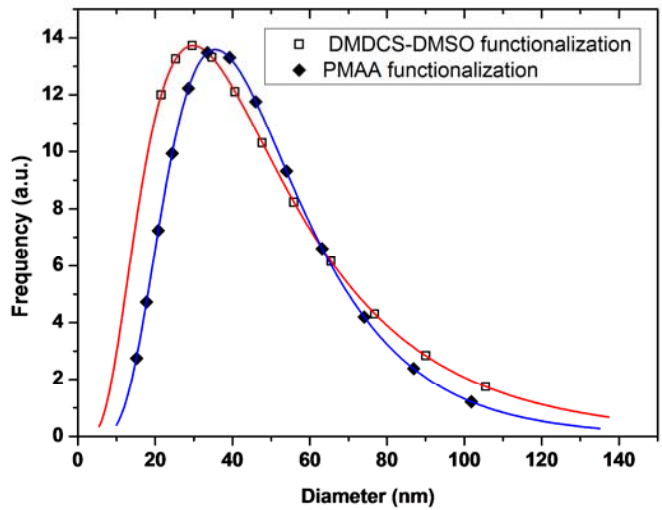 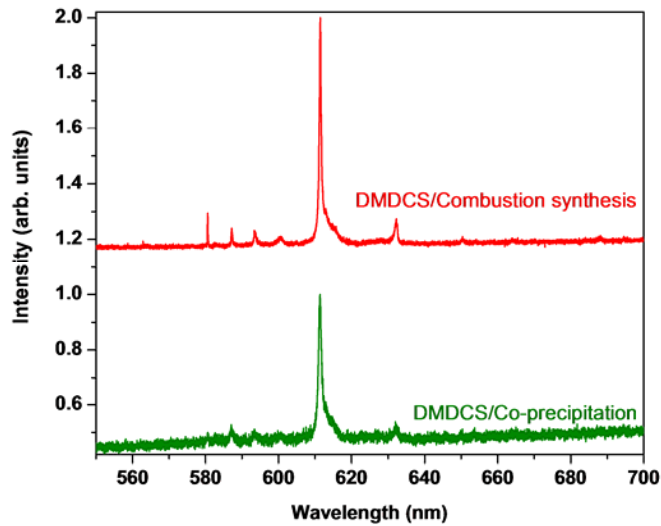

Figure 6. Particle size distribution of dispersed Eu:$Lu_2O_3$ nanopowders using different coupling agents by dynamic light scattering. The fitting of lognormal line profiles is superimposed.

Figure 7. Fluorescence spectra of DMDCS-DMSO surface functionalized co-precipitated Eu:$Lu_2O_3$ nanoparticles (bottom) and DMDCS-DMSO surface functionalized combustion-synthesized Eu:$Lu_2O_3$ nanoparticles (top).

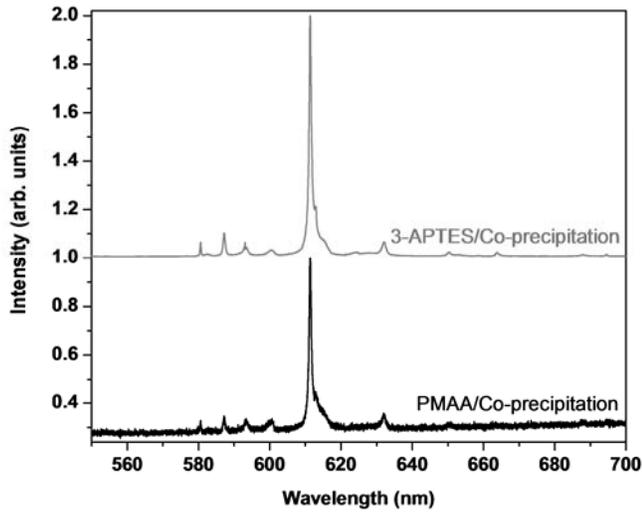

Figure 8. Fluorescence spectra of 3-APTES surface functionalized co-precipitated Eu:Lu$_2$O$_3$ nanoparticles (top) and of PMAA coated co-precipitated Eu:Lu$_2$O$_3$ nanoparticles (bottom)

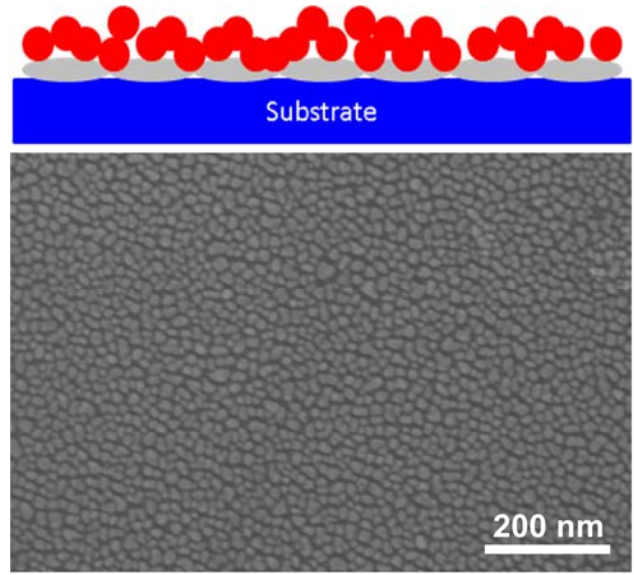

Figure 9. Thin film design (top) indicating silver nanoparticles in grey and Eu:Lu$_2$O$_3$ nanoparticles in red. SEM of silver nanoparticles (bottom).

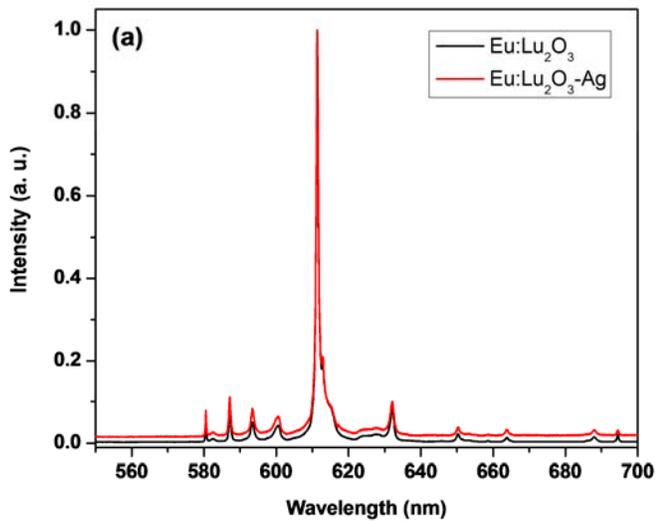
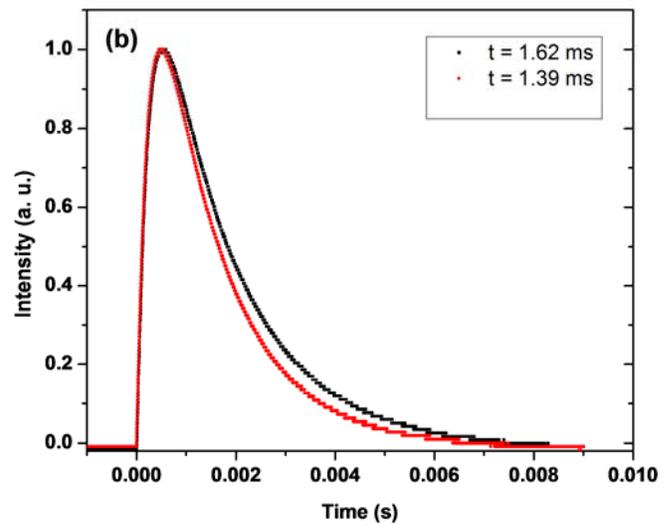

Figure 10. Comparison of fluorescence (a) and fluorescence lifetime (b) of Eu:Lu$_2$O$_3$ nanoparticles (black) and Eu:Lu$_2$O$_3$ nanoparticles on top of silver nanoparticles (red). The reduction in fluorescence lifetime is 14%.